\documentclass[10pt,conference]{IEEEtran}
\IEEEoverridecommandlockouts
\usepackage{amsmath,amssymb,amsfonts}
\usepackage{comment}
\usepackage{cite}
\usepackage{graphicx}
\usepackage{textcomp}
\usepackage[breaklinks,colorlinks,linkcolor=blue,citecolor=blue,urlcolor=blue]{hyperref}
\usepackage{colortbl}
\usepackage{multirow}
\usepackage{footnote}
\usepackage{tablefootnote}
\usepackage{float}
\usepackage{threeparttable}
\usepackage{pifont}
\usepackage{textcomp}
\usepackage{tabularx}
\usepackage[table]{xcolor}

\usepackage{algorithmic}
\usepackage[ruled,vlined,linesnumbered,commentsnumbered]{algorithm2e}
\usepackage{arydshln}

\def\BibTeX{{\rm B\kern-.05em{\sc i\kern-.025em b}\kern-.08em
    T\kern-.1667em\lower.7ex\hbox{E}\kern-.125emX}}

\begin{document}

\newcommand{\ml}[1]{{\color{red}\bf [Meng: #1]}}
\newcommand{\yx}[1]{{\color{blue}\bf [Yixuan: #1]}}
\newcommand{\wrj}[1]{{\color{green}\bf [wrj: #1]}}
\newcommand{\xt}[1]{{\color{orange}\bf [xt: #1]}}

\newcommand{\red}[1]{{\color{red}\bf (#1)}}
\newcommand{\method}{ASCEND}

\title{\method: Accurate yet Efficient End-to-End Stochastic Computing Acceleration of Vision Transformer}

\author{
    Tong Xie$^{1\dag}$,
    Yixuan Hu$^{1\dag}$,
    Renjie Wei$^{1\dag}$,
    Meng Li$^{213*}$,
    Yuan Wang$^{13}$, Runsheng Wang$^{134}$, and Ru Huang$^{134}$
\\
\textit{$^1$School of Integrated Circuits \& $^2$Institute for Artificial Intelligence, Peking University, Beijing, China} \\
\textit{$^3$Beijing Advanced Innovation Center for Integrated Circuits, Beijing, China} \\
\textit{$^4$Institute of Electronic Design Automation, Peking University, Wuxi, China} \\
$^\dag$Equal contribution
$^*$Corresponding author: meng.li@pku.edu.cn
\vspace{-15pt}
\thanks{
This work was supported in part by the National Key R$\&$D Program of China (2020YFB2205502), NSFC (62125401) and the 111 Project (B18001). }
}
\maketitle
\begin{abstract}




  Stochastic computing (SC) has emerged as a promising computing paradigm for neural acceleration.
  However, how to accelerate the state-of-the-art Vision Transformer (ViT) with SC remains unclear.
  Unlike convolutional neural networks, ViTs introduce notable compatibility and efficiency
  challenges because of their nonlinear functions, e.g., softmax and Gaussian Error Linear Units (GELU).
  In this paper, for the first time, a ViT accelerator based on end-to-end SC, dubbed \method, is proposed.
  \method~co-designs the SC circuits and ViT networks to enable accurate yet efficient acceleration. 
  To overcome the compatibility challenges,
  \method~proposes a novel deterministic SC block for GELU and leverages an SC-friendly iterative approximate algorithm to design an accurate and efficient softmax circuit.
  To improve inference efficiency, \method~develops a two-stage training pipeline to produce accurate
  low-precision ViTs.
  With extensive experiments, we show the proposed GELU and softmax blocks achieve 56.3\%
  and 22.6\% error reduction compared to existing SC designs, respectively, and reduce
  the area-delay product (ADP) by 5.29$\times$ and 12.6$\times$, respectively.
  Moreover, compared to the baseline low-precision ViTs, \method~also achieves significant
  accuracy improvements on CIFAR10 and CIFAR100.
  

\end{abstract}

\vspace{-5pt}
\section{Introduction}\vspace{-2pt}
\label{sec:intro}
Transformer has been widely applied to computer vision tasks, 
including classification \cite{dosovitskiy2020image}, object detection \cite{carion2020end},
segmentation \cite{gu2022multi}, etc. With the high model capacity and large
receptive field \cite{dosovitskiy2020image}, vision transformers (ViTs) have achieved
state-of-the-art (SOTA) accuracy compared to convolutional neural networks
(CNNs). However, the high accuracy is achieved at the cost 
of a rapid increase in ViT parameters and computation \cite{liu2021swin},
which calls for more efficient neural accelerators.


Stochastic computing (SC) emerges as a new computing paradigm and has attracted
much attention for neural acceleration in recent years 
\cite{zhang2020sorting,kim2016dynamic,ren2017scdcnn, li2017towards,li2018heif,zhang2020accurate}. 
By representing a number with a stochastic bitstream, in which
the probability of 1's denotes the value, 
SC achieves simplified arithmetic logics and improved fault tolerance \cite{li2011using}.
End-to-end SC avoids the back-and-forth conversion between the binary and
the SC representations, achieving even higher hardware 
efficiency \cite{zhang2020sorting, zhang2020accurate}.

\begin{figure}[!tb]
 \centering
 \includegraphics[width=0.90\linewidth]{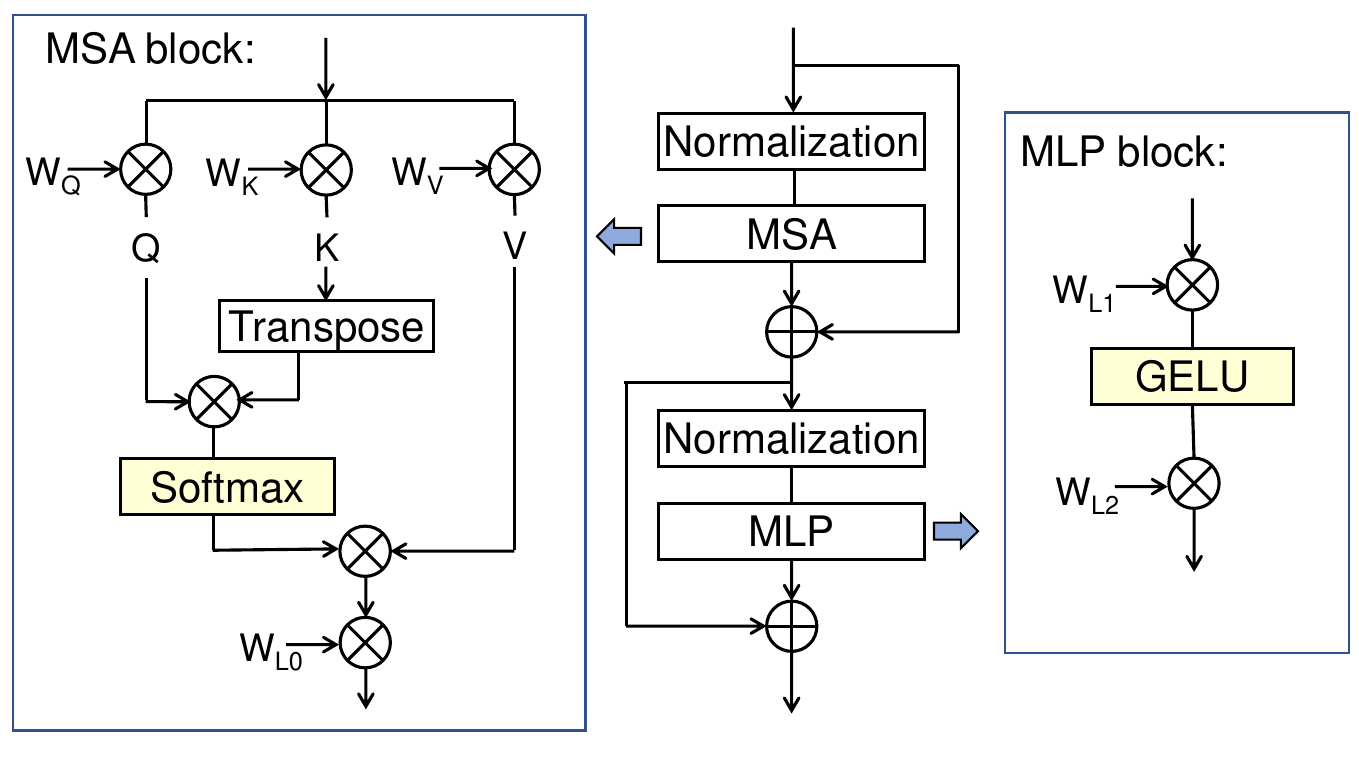}
\vspace{-14pt}
 \caption{The diagram of a transformer encoder block.}
 \vspace{-17pt}
 \label{fig:vit_arch}
\end{figure}

However, we observe end-to-end SC encounters major compatibility and efficiency challenges for ViT acceleration. 
On one hand, ViTs have more complex nonlinear functions such as Gaussian Error Linear Unit (GELU) in the multi-head self-attention (MSA) block and softmax in the multi-layer perceptron (MLP) block \cite{dosovitskiy2020image} 
as shown in Fig.~\ref{fig:vit_arch}.
While these functions are important to the accuracy of ViTs \cite{zeng2022mpcvit}, 
their SC implementations are not well studied in existing works.
On the other hand, ViTs can be harder to quantize and require higher computation precision \cite{bai2020binarybert,liu2022bit},
resulting in an exponential increase of SC representation bitstream length
(BSL) \cite{hu2023accurate} and drastic degradation of inference efficiency.

Therefore, a natural question to ask is \textit{whether it is possible to
leverage SC to realize accurate yet efficient ViT acceleration.} 
In this paper, we provide an affirmative answer and propose \method, the first end-to-end SC-based ViT accelerator.
\method~features a co-design of the SC circuits and ViT networks to address the aforementioned challenges.
At the circuit level, we observe implementing the division and exponential
functions in GELU and softmax directly is challenging for SC. 
Therefore, we propose a novel deterministic SC block for the GELU function. 
We leverage an SC-friendly iterative approximate algorithm and implement the corresponding SC circuits to achieve an accurate and efficient softmax function.
At the network level, a two-stage pipeline is proposed for SC-friendly ViT training,
which improves the accuracy of low-precision ViTs through progressive quantization and SC circuit-aware fine-tuning.
Our contributions can be summarized as follows:
\begin{itemize}
  \item We propose the first end-to-end SC accelerator for ViT and design novel SC blocks for GELU and softmax.
  \item We propose a two-stage training pipeline for SC-friendly low-precision ViT, which features progressive quantization and SC circuit-aware fine-tuning.
  \item We demonstrate 56.3\% and 22.6\% mean average error (MAE) reduction
    compared to existing baselines for GELU and softmax, respectively while achieving 5.29$\times$
    and 12.6$\times$ area-delay product (ADP) reduction. \method~also achieves significant accuracy
    improvement compared to baseline low-precision ViTs.

  
\end{itemize}




\section{Background}
\label{sec:background}

\vspace{-1pt}
\subsection{SC Overview}
\label{subsec: background_SC}
\vspace{-1pt}
Unlike traditional binary encoding, each bit in an SC representation carries equal weight. 
It leverages the probability of 1's in the bitstream to represent 
values. 
Since probabilities inherently lie within $[0, 1]$, SC encoding relies on scaling factors to map the probability
to a desired range.

The simplest unipolar encoding expresses the value of $[0,1]$ with the probability of 1's, denoted as $p$, directly.
To represent both positive and negative values, bipolar encoding defines the value as $2p-1$, thus mapping the probability to the range of $[-1,1]$.
For these encoding schemes, multiplications and additions can be
implemented with simple logics, e.g., AND and MUX gates for unipolar encoding \cite{ren2017scdcnn}.
However, the stochastic nature of these encoding schemes usually leads to large computation errors and fluctuation \cite{zhang2020accurate}.

Another encoding format known as thermometer encoding \cite{zhang2020accurate} offers a deterministic approach where all the 1’s appear at the beginning of the bitstream. A data $x$ is represented with an $L$-bit sequence as $x = \alpha x_q$, where $\alpha$ is the scaling factor and $x_q = \sum\limits_{i=0}^{L-1}x[i]-\frac{L}{2}\in[-\frac{L}{2},\frac{L}{2}]$. 
The multiplication can thus be implemented based on a truth table \cite{zhang2020accurate}. The addition can be realized by concatenating the input bitstreams together using a Bitonic Sorting Network (BSN) \cite{zhang2020sorting}.

\begin{table}[!tb]
    \centering
        \caption{Gap between model requirements and existing SC circuits.}
     \vspace{-8pt}
        \label{tab: main table}\setlength{\tabcolsep}{3pt}
    \scalebox{0.95}{
    \begin{tabular}{c|c|c|c|c}
        \hline \hline
        SC&Supported&Encoding&Supported&Implementation \\
        Design&Model&Format&Function&Method \\
        \hline \hline
        \cite{kim2016dynamic,li2017towards,ren2017scdcnn} & CNN & stochastic  & tanh, sigmoid & FSM\\
        
        \cite{li2018heif} & CNN & stochastic  & ReLU & FSM \\
        
        \cite{yuan2017softmax,hu2018efficient} & CNN & stochastic  & softmax & FSM, binary units \\
       
        \cite{zhang2020sorting,hu2023accurate} & CNN & deterministic  & ReLU& SI\\
\rowcolor{lightgray}    Ours & ViT & deterministic  & GELU, softmax & Gate-Assis. SI, BSN\\
        \hline \hline
    \end{tabular}
    }
     \vspace{-12pt}
\end{table}

\vspace{-2pt}
\subsection{SC-based Method for Nonlinear Functions}
\label{subsec:nonlinear}
\vspace{-1pt}
There are three categories of SC designs for nonlinear functions. The first category is based on finite state machines (FSMs) and saturated counters. By adjusting the FSM transition conditions, different nonlinear functions can be implemented~\cite{kim2016dynamic,li2017towards,ren2017scdcnn,li2018heif}. However, due to the nature of sequential processing, the design often requires very long bitstreams while the output error is still large.

The second category relies on the Bernstein polynomial to approximate nonlinear functions \cite{qian2011uniform}. This algorithm requires high degree polynomials which have many terms and long input bitstreams to reduce the approximation error. Moreover, a large number of stochastic number generators (SNGs) are needed, resulting in a very high hardware cost.

The third category, namely selective interconnect (SI), is proposed for thermometer coding. SI processes the whole input bitstream in parallel and computes nonlinear functions by controlling the position of output transitions, which enables accurate computation with a relatively short bitstream \cite{hu2023accurate,zhang2020sorting}. However, it can only support monotonic functions. 

There are many works employing these methods for nonlinear functions in CNNs,
but none of them can support GELU and softmax accurately, as shown in Table \ref{tab: main table}. \cite{kim2016dynamic,li2017towards,li2018heif,ren2017scdcnn} design FSMs for tanh,
sigmoid, and ReLU in traditional
stochastic encoding format. But they all
need very long bitstreams to reduce the computation fluctuations. \cite{yuan2017softmax,hu2018efficient} implements softmax for the last layer of CNNs with various binary compute units and multiple SC-binary conversions. However, only the relative order of outputs is preserved while the computed values still exhibit a large error.
\cite{zhang2020sorting,hu2023accurate} are representative end-to-end SC architectures for CNNs employing deterministic thermometer encoding format, but only monotonic nonlinear functions, e.g., ReLU, sigmoid, can be supported.



\vspace{-2pt}
\section{Challenges of SC-based ViT Acceleration}
\label{sec:challenge}
\vspace{-2pt}
In this section, we discuss the challenges of designing SC blocks for GELU and softmax.
\vspace{-5pt}
\subsection{SC Blocks for GELU}
\vspace{-2pt}

\begin{figure}[!tb]
    \centering
    \includegraphics[width=1\linewidth]{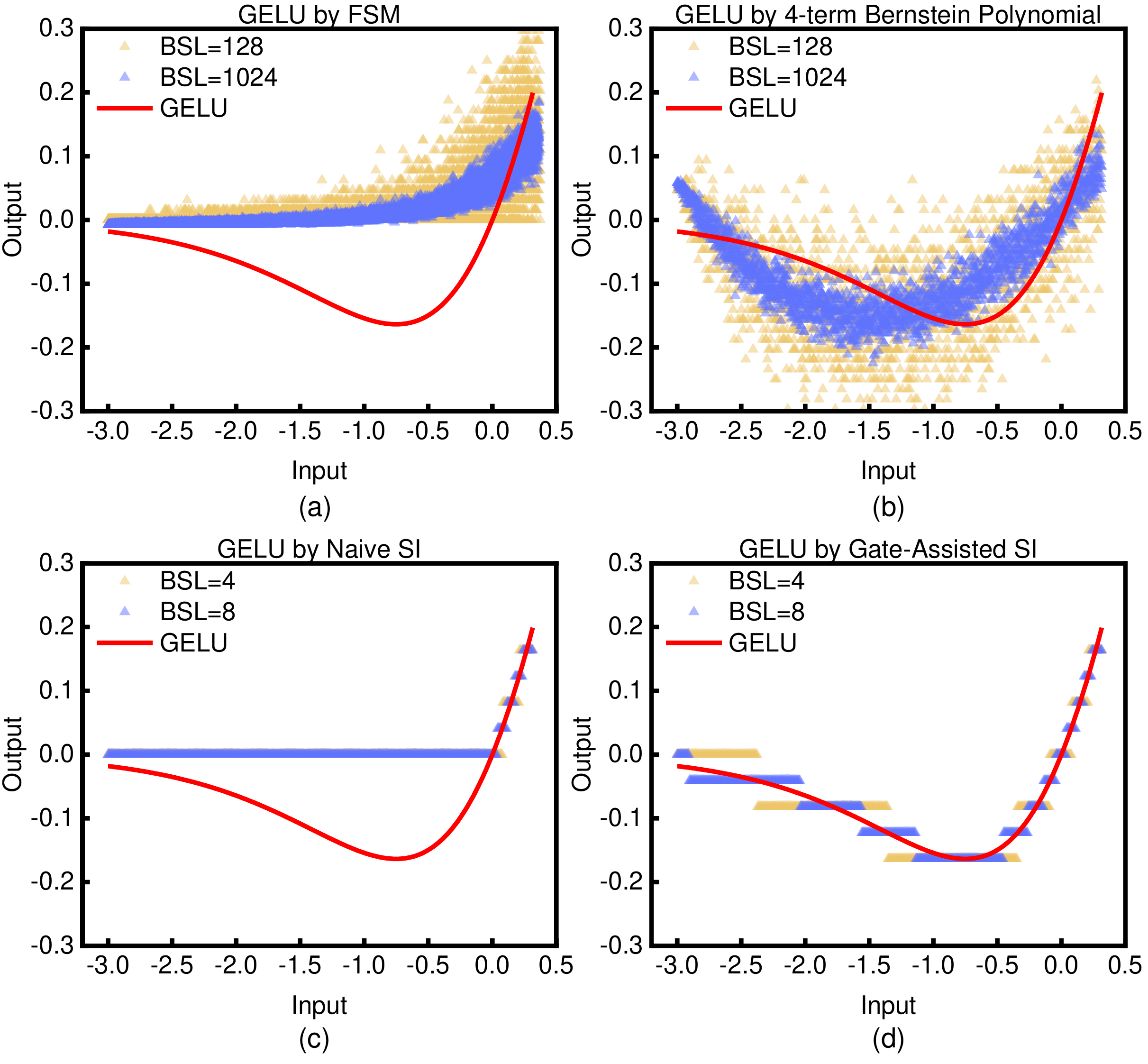} \vspace{-20pt}
    \caption{GELU by (a) FSM-based design, (b) 4-term Bernstein polynomial , (c) naive SI-based design, and (d) the proposed gate-assisted SI design. }
    \label{fig: scatter}
    \vspace{-10pt}
\end{figure}

Existing methods introduced in Section~\ref{subsec:nonlinear} all suffer from different
limitations when implementing GELU:

\emph{FSM-based methods} struggle with implementing GELU for both positive and negative input ranges.
As shown in Fig.~\ref{fig: scatter} (a), when the input is a small negative value,
the output of FSM-based methods saturates at 0, resulting in systematic errors \cite{li2018heif}. For the positive range, they are forced to handle a very long BSL to reduce random errors, significantly impacting hardware efficiency \cite{hu2023accurate} .


\emph{Bernstein polynomial-based methods} approximate nonlinear functions through polynomial fitting. 
Their hardware cost increases with the degree  of polynomials.
As shown in Fig. \ref{fig: scatter} (b), low-degree Bernstein polynomial fitting
suffers from a high computation error while high-degree fitting results in a drastic increase in 
hardware cost. 
Moreover, this design also exhibits noticeable computation fluctuations, which forces to use long bitstreams and further increases the inference cost.

\emph{SI-based methods} can only implement monotonic functions while GELU is non-monotonic \cite{zhang2020sorting}.
As shown in Fig.~\ref{fig: scatter} (c), naively adopting SI-based methods suffers from a high
computation error for the negative input range. However, it should be noted the positive range
can be accurately implemented even for short bitstreams.

\vspace{-2pt}
\subsection{SC Blocks for Softmax}
\label{subsec: challenge of softmax}
\vspace{-2pt}

Softmax involves computing the division and exponential functions,
both of which are challenging to support with SC.
The exponential function usually demands high precisions for its
input and output, leading to an exponential increase in operands' BSLs \cite{hu2023accurate}.
The SC-based division function incurs a high hardware cost
due to extensive usage of SNGs with JK flip-flops \cite{naderi2011delayed} or adaptive digital elements \cite{chen2016design} or comparators \cite{chen2016design}. It also suffers from high computation error as
only approximate divisions like ${p_x}/(p_x+p_y)$ can be implemented
rather than $p_x/p_y$ \cite{chen2016design,canals2015new}.
Moreover, existing SC implementations for division and exponential
function all leverage sequential bitstream processing with
high random fluctuations.

\vspace{-6pt}
\subsection{Efficiency}
\vspace{-3pt}

Apart from the compatibility of nonlinear functions, SC-based 
ViT accelerators also encounter efficiency bottlenecks.
Compared to CNNs, ViT often requires significantly larger
computational resources and the problem is further exacerbated by the coding efficiency of SC. 
As mentioned in Section~\ref{subsec: background_SC},
a bitstream of $L$b BSL can only represent $L+1$ values.
Hence, data of $n$-bit binary precision require
$2^n$b BSL to represent in SC, indicating an exponential increase
of BSL. In parallel SC circuits, higher
precision results in an exponential increase in area \cite{hu2023accurate}, 
whereas
in serial SC circuits, it signifies an exponential growth in delay
\cite{ren2017scdcnn}.
This trade-off becomes even more crucial for ViT, given
its already demanding computational resource requirements.

\begin{figure}[!tb]
 \centering
 \includegraphics[width=0.9\linewidth]{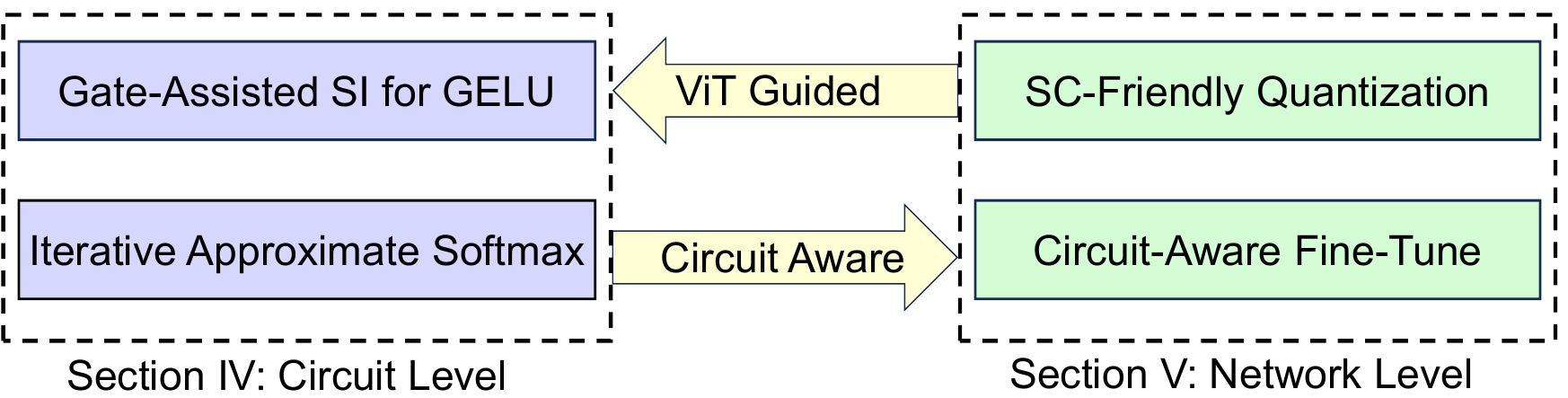}
\vspace{-10pt}
 \caption{The circuit/network co-design of the proposed ASCEND.}
 \vspace{-15pt}
 \label{fig:fig_method}
\end{figure}
\section{Efficient SC-based ViT Accelerator Circuit}
\label{sec:circuit}

To address the challenges in Section \ref{sec:challenge}, we propose the ASCEND framework for co-designing the SC circuits in this section and ViT networks in Section \ref{sec:network}. The overall flow is illustrated in Fig. \ref{fig:fig_method}.

\vspace{-2pt}
\subsection{Gate-Assisted SI for GELU Activation}
\vspace{-3pt}

Since the three methods mentioned above all have their limitations, we propose gate-assisted SI to implement the GELU function.

\begin{figure}[!tb]
    \centering
    \includegraphics[width=0.82\linewidth]{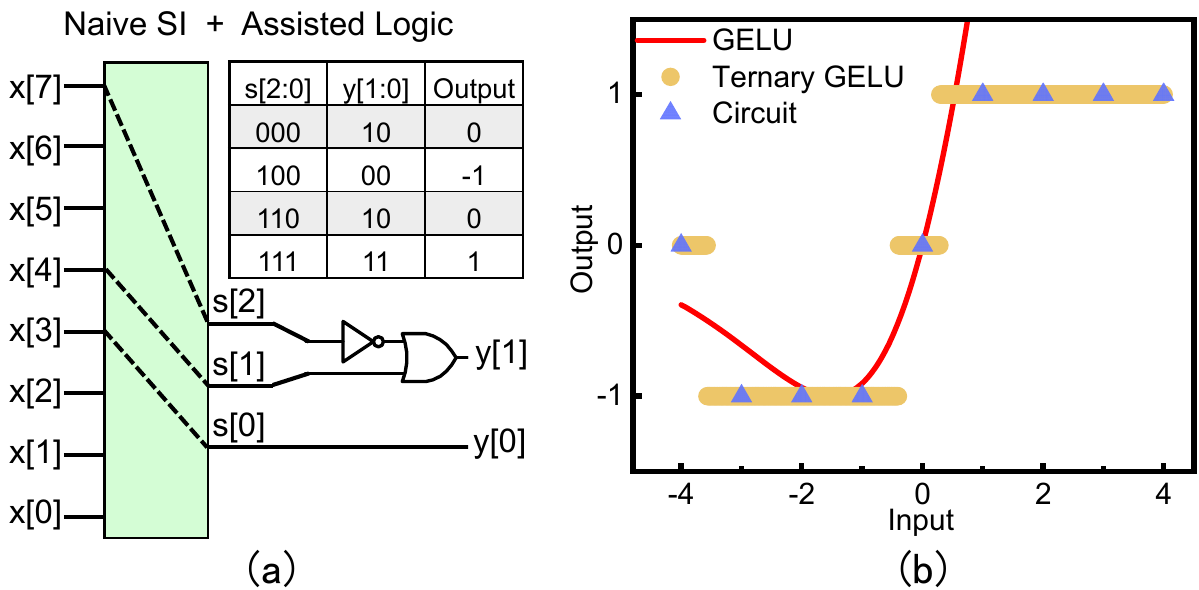} 
    \vspace{-5pt}
    \caption{(a) Gate-assisted SI implements non-monotonic functions with the help of simple combinational logics. (b) Ternary GELU implemented by (a).}
    \label{fig: SI GELU}
    \vspace{-15pt}
\end{figure}

SI-based designs leverage parallel bitstreams and can obtain all the input information for the output bit, allowing for accurate results. 
But naive SI directly outputs selected bits, which increases the number of 1's in the output as the number of 1's in the input grows, limiting it to monotonic functions.
To address this issue, we propose gate-assisted SI, which introduces extra gates and outputs the logical results of selected bits rather than directly outputting them, as shown in Fig.~\ref{fig: SI GELU} (a). 
For example, a NOT gate and an AND gate can assist in implementing non-linear functions like GELU, which exhibit a decrease followed by an increase.

In the case of ternary GELU with an 8b BSL input and a 2b BSL output, as illustrated in Fig. \ref{fig: SI GELU}, we use assisted logic to set the outputs as $y[1]$ = !$s[2]$ \& $s[1]$, $y[0]$ = $s[0]$.
The selected bits are controlled by selection signals derived from the input. When the input is small, all the selected bits are 0, resulting in the output $y[1:0]$ being "10" corresponding to the value 0. As the input increases, $x[7]$ to $x[0]$ gradually transitions from 0 to 1, and so do the selected bits.
When $s[2]$ becomes 1, the output $y[1:0]$ changes to "00" corresponding to a decrease to -1. When $s[1]$ also becomes 1, the output $y[1:0]$ reverts to "10" corresponding to an increase back to 0. When the input increases further and $s[0]$ becomes 1, the output $y[1:0]$ changes to "11" corresponding to an increase to 1. 

Fig. \ref{fig: scatter} (d) illustrates the GELU functions implemented by gate-assisted SI with different precision.
The proposed GELU design is free of random fluctuations, allowing the exact implementation of the required GELU function. 

\vspace{-5pt}
\subsection{SC Block of Iterative Approximate Softmax}
\vspace{-2pt}

Due to the limitations mentioned in Section~\ref{subsec: challenge of softmax}, existing research has not been able to  accurately implement an efficient SC-based softmax. We propose introducing an iterative approximation of softmax \cite{zhang2021AxCsoftmax} to address this issue, as shown in Algorithm \ref{algo:iterative_softmax}. 
With an $m$-dimensional parameterized function $y(t)=\text{softmax}(tx)$, where $y(0)=1/m$ and $y(1)=\text{softmax}(x)$, we can approximate $y(1)$, i.e., the softmax, from $y(0)$ using a summation of $k$ terms instead of an integral:
$$
y(1)=y(0)+\int_0^1y'(t)\mathrm dt\approx y(0)+\sum\limits_{j=0}^{k-1} y'(\dfrac{j}{k})\cdot\dfrac{1}{k}
$$

Due to the exponential term in $y(t)$, we compute and observe that $y'(t)$ can be straightforwardly expressed based on the value of $y(t)$. Therefore, we can iteratively calculate and obtain $y'(0), y(1/k), y'(1/k), y(2/k), \dots, y(1)$, starting from the known value of $y(0) = 1/m$.

\begin{algorithm}[!tb]
    \caption{Iterative Approximation of Softmax}
    \label{algo:iterative_softmax}
    \SetKwProg{Fn}{Function}{:}{end}
    \SetKwComment{Comment}{/* }{ */}
    \SetKwInOut{Input}{Input}
    \SetKwInOut{Output}{Output}
    \Input{Softmax input vector $x$ of dimension $m$}
    \Output{Vector $y$ after $k$ iterations}
    Initialize $y^0_i = \frac{1}{m} \quad \forall i \in \{0, 1, \ldots, m-1\}$\;
    \For{$j=1:k$}{
    $z_i = x_i \cdot y^{j-1}_i \quad \forall i \in \{0, 1, \ldots, m-1\}$\;
    $y^j_i = y^{j-1}_i+ [z_i -y^{j-1}_i \cdot \mathrm{sum}(z)] / k  \quad \forall i \in \{0, 1, \ldots, m-1\}$\;
    }   
    \KwRet $y$\; 

\end{algorithm}

This iterative approximation simplifies the intricate computations of an $m$-dimensional vector $x$ into iterations of $t$, avoiding the challenges of implementing division, exponentiation, and higher-order multiplications between bitstreams in SC.
Instead, only multiplication and accumulation operations are required, along with the division of bitstreams by a constant $k$, which enables the efficient softmax circuits in SC.



\begin{figure}[!tb]
    \vspace{-10pt}
    \centering
    \includegraphics[width=1\linewidth]{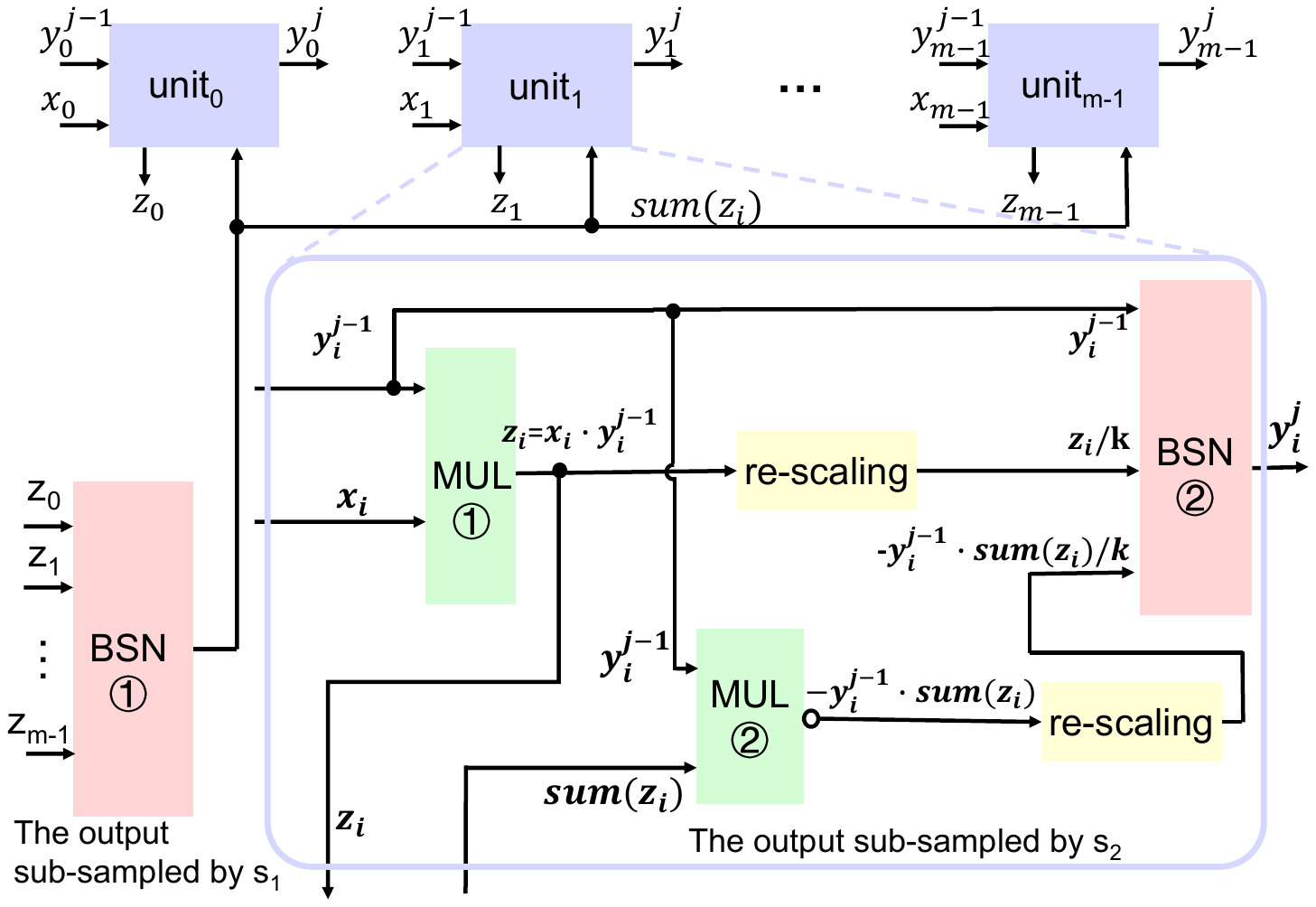}
    \vspace{-15pt}
    \caption{SC circuit block of Iterative Approximate Softmax.}
    \label{fig: softmax circuit}
    \vspace{-8pt}
\end{figure}

Based on Algorithm \ref{algo:iterative_softmax}, we propose the corresponding SC circuit in Fig.~\ref{fig: softmax circuit}. 
Here, multiplication and accumulation can be implemented as mentioned in Section \ref{subsec: background_SC}, and division by a constant $k$ can be implemented by just dividing the scaling factor by $k$ without any operation on the bitstream.

The circuit block consists of $m$ compute units for  $m$ elements of the softmax row vector and a global BSN \ding{172}.
Each compute unit takes inputs $x_i$ and $y^{j-1}_i$ and computes $y^j_i$ for the current iteration.
 Multiplier \ding{172} executes the multiplication for obtaining $z_i$ in the 3rd line of Algorithm \ref{algo:iterative_softmax}. 
Multiplier \ding{173} performs the multiplication $y^{j-1}_i \cdot \mathrm{sum}(z)$. 
Meanwhile, BSN \ding{172} calculates the summation of $z_i$.


Up to this point, we have acquired all terms in the 4th line of Algorithm \ref{algo:iterative_softmax}. We utilize two re-scaling blocks \cite{hu2023accurate} to align their scaling factors. And BSN \ding{173} carries out the final accumulation to obtain $y^j_i$,
which yields $y^j_i = y^{j-1}_i+ [z_i -y^{j-1}_i \cdot \mathrm{sum}(z)] / k$ as the output of this iteration.
The final result of softmax is available after $k$ iterations. Table~\ref{tab:paramters} lists the parameters used to describe the detail of the circuit.

\begin{table}[!tb]
    \centering
    \vspace{-15pt}
    \caption{Parameters in the proposed softmax circuit block.} 
    \vspace{-6pt}
    \label{tab:paramters}
    \begin{tabular}{c|c}
        \hline \hline
         Symbol     & Meaning\\ 
        \hline \hline 
        $m$     & length of the row vector  \\ 
        \hline    
        $k$     & count of iteration      \\   
        \hline 
        $B_x$   & bitstream length of  $x$  \\   
        \hline 
        $\alpha_x$& scaling factor of $x$ \\     
        \hline 
        $B_y$   & bitstream length of $y$      \\
        \hline 
        $\alpha_y$& scaling factor of $y$\\     
        \hline 
        $s_1$   & sub-sample rate of $\text{sum}(z)$  \\ 
        \hline      
        $s_2$   & sub-sample rate of $\text{sum}(z) \times y$  \\ 
        \hline \hline
        \end{tabular}
        \vspace{-20pt}
\end{table}

Furthermore, \textit{we emphasize the general SC-friendly nature of iterative approximate softmax and it can be also applied to other SC designs.}

\section{SC-friendly Low-precision ViT}
\label{sec:network}

\begin{figure}[!tb]
\centering
\includegraphics[width=0.85\linewidth]{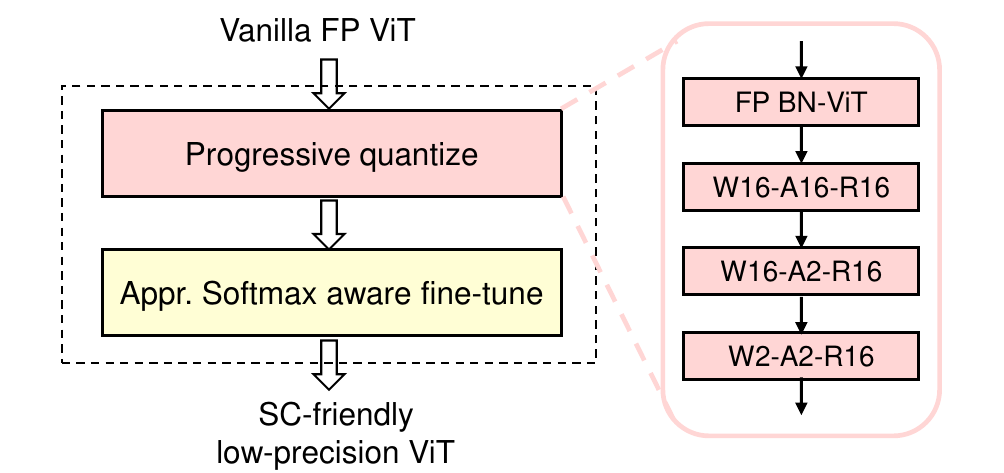}
\vspace{-5pt}
\caption{The proposed SC-friendly low-precision ViT training pipeline.}
\label{fig: pipeline}
\vspace{-10pt}
\end{figure}

The circuit design mentioned above solves the challenges posed by GELU and softmax in ViT.
To further enhance ViT's compatibility with SC, we substitute layer normalization (LN) with batch normalization (BN). 
 With knowledge distillation (KD), the replacement leads to less than 0.1\% accuracy impact on CIFAR10 and CIFAR100 datasets.
To address the challenge in efficiency, we quantize weight and activation to 2b BSL, and residual to 16b BSL, denoted as W2-A2-R16, following \cite{hu2023accurate}.
However, we find that direct quantization to low-precision ViT leads to a severe accuracy drop as shown in Table \ref{tab:network compare}.
Therefore, we propose a two-stage training pipeline for SC-friendly low-precision ViT as shown in Fig.~\ref{fig: pipeline}.


The first stage is the progressive quantization.
Inspired by \cite{martinez2020training,liu2022bit}, we start from the full-precision (FP) model and use a three-step procedure, i.e., FP $\rightarrow$ W16-A16-R16 $\rightarrow$ W16-A2-R16 $\rightarrow$ W2-A2-R16 to achieve the low-precision model.
In each step, we use the output from the last step as initialization.
We also use KD to guide the training of low-precision model.
We use the FP model as the teacher of the first step, and for the last two steps, we use W16-A16-R16 as the teacher, 
which is closer to the resulting model and provides sufficient information for the student to learn.
The KD objective is 
\vspace{-3pt}
\begin{equation*}
Loss=\ell_{KL}(Z_{s}, Z_{t}) + \beta \cdot \frac{1}{M} \sum_{i=1}^{M} \ell_{MSE}(S_{i}, T_{i})
\end{equation*}
where $\ell_{KL}$ and $\ell_{MSE}$ are the Kullback–Leibler (KL) divergence loss and mean squared error (MSE) loss, respectively.
$Z_{s}$ and  $Z_{t}$ denote the logits of the teacher and student model respectively. 
$S_{i}$ and $T_{i}$ denote the output of the $i$-th layer of the student model and teacher model, with a total of $M$ layers.
$\beta$ is the coefficient balancing the KL loss and the MSE loss, which is set to 2 experimentally. 

The second stage is approximate softmax aware fine-tuning.
Considering the hardware efficiency, we replace the softmax in the W2-A2-R16 ViT with the iterative approximate softmax.
However, this brings an accuracy degradation to the model.
Thus we fine-tune the model after the replacement to adapt it to the approximate softmax, which will make up for the accuracy loss.
So far, we have obtained SC-friendly low-precision ViT.

\section{Experiment}
\label{sec:experiment}
\vspace{-1pt}
\subsection{Experiment Setup}
\vspace{-1pt}
\emph{Hardware Evaluation}: we implement the register transfer level (RTL) code of our SC designs and existing SC designs 
and then, synthesize them with Synopsys Design Compiler using TSMC 28nm technology library. 
We report the hardware metrics based on the synthesis results.

We evaluate different SC designs in terms of area, delay, and computation error.
We also calculate the area-delay product (ADP) for hardware cost comparison. 
To evaluate computation errors, we collect the input vectors of softmax for each layer in ViT and generate test vectors sampled from the overall distribution. 
We calculate the mean average error (MAE) between the SC circuit outputs and the correct results.

\emph{Network Evaluation}:
we evaluate the proposed training method on a lightweight ViT, which has 7 layers and 4 heads on CIFAR10 and CIFAR100 datasets following \cite{hassani2021escaping}.
We use learned step size quantization (LSQ) to quantize both weights and activations \cite{esser2019learned}.
For the training settings, we use AdamW optimizer \cite{loshchilov2017decoupled} with a momentum of 0.9. 
We set the batch size to 128. 
For the first stage in the training pipeline, we train the model for 300 epochs with an initial learning rate of 7.5$\times 10^{-4}$.
For the second stage, we replace the softmax with iterative approximate softmax and fine-tune the model for 30 epochs with an initial learning rate of 5$\times 10^{-6}$. 

\vspace{-5pt}
\subsection{Main Results}
\vspace{-1pt}

\subsubsection{SC Circuits Comparison} 
We evaluate our proposed blocks for the GELU and softmax function. The design space for different approximate configurations of iterative approximate softmax is also explored.

\emph{GELU Block Comparison} 
We focus on comparing the proposed gate-assisted SI with Bernstein polynomial-based
design since it is the only baseline that can implement GELU in the 
negative input range as described in Section~\ref{subsec:nonlinear}.

\begin{figure}[!tb]
 \centering
 \vspace{-8pt}
 \includegraphics[width=1\linewidth]{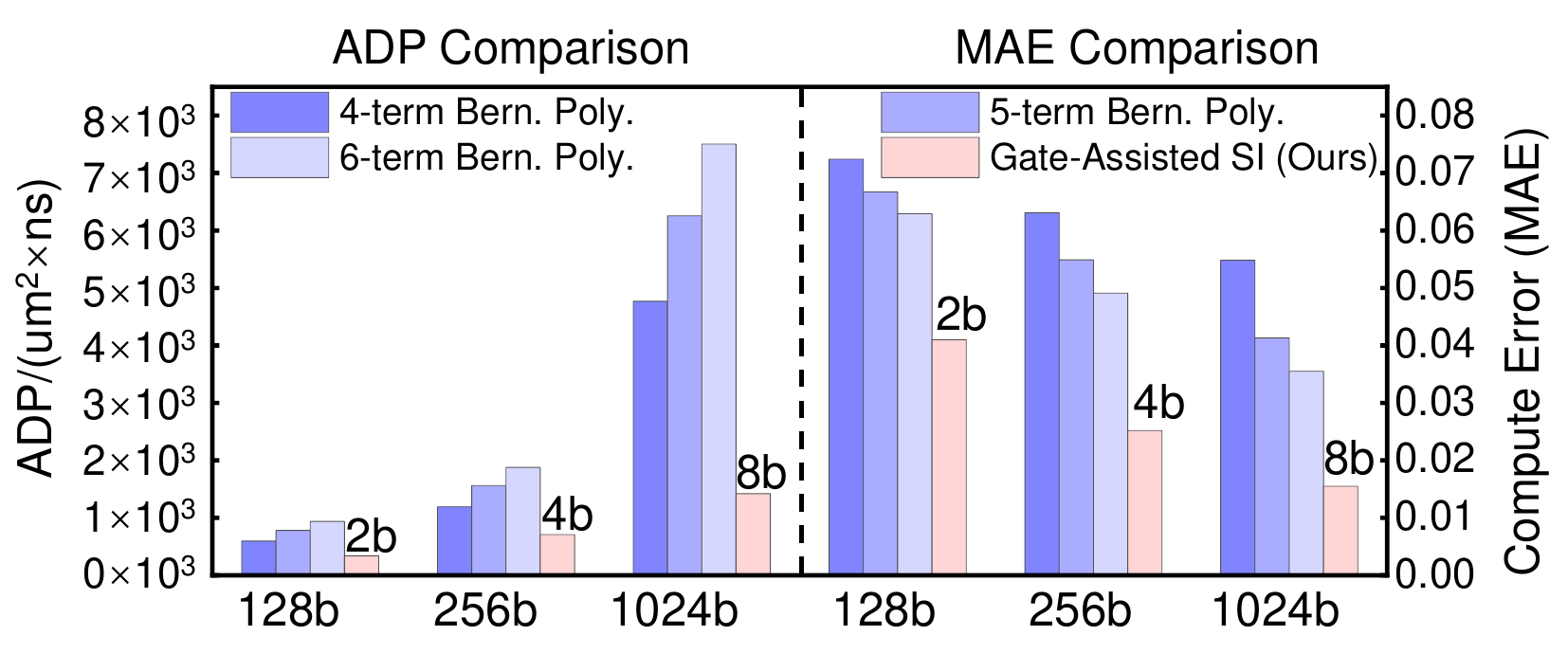}
 \vspace{-25pt}
 \caption{GELU block comparison with different BSLs.}
 \label{fig: GELU compare}
\end{figure}

While the Bernstein polynomial-based design requires numerous cycles for sequential computation, our design is fully parallel, enabling significant
latency reduction. 
As shown in Table \ref{tab: GELU compare}, the 8b BSL gate-assisted SI achieves ADP reduction from 3.36$\times$ to 5.29$\times$ compared to the 1024b BSL baseline method with different polynomial terms.
At the same time, we also reduce computation errors by 71.7\% to 56.3\%. 
If a larger computational error is allowed, we can further reduce the ADP by 4.15$\times$ from 1420 to 342 $\rm um^2 \cdot us$.
More ADP and MAE comparisons are also visualized in Fig. \ref{fig: GELU compare}. 

\begin{table}[!tb]
\centering
\caption{Compare area, delay, area-delay product (ADP), and mean average error (MAE) 
across different SC circuits for GELU.}
\vspace{-8pt}
\label{tab: GELU compare} 
\setlength{\tabcolsep}{5pt}
\begin{tabular}{cc|cccc}
\hline \hline
\multicolumn{2}{c|}{Design} & \begin{tabular}[c]{@{}c@{}}Area$\downarrow$ \\ (um²)\end{tabular} & \begin{tabular}[c]{@{}c@{}}Delay$\downarrow$ \\ (ns)\end{tabular} & \begin{tabular}[c]{@{}c@{}}ADP$\downarrow$\\(um²·ns)\end{tabular} & MAE$\downarrow$ \\ \hline \hline
Bernstein  & 4-term poly& 58.2 & 81.92 & 4769 & 0.0548    \\ 
 polynomial& 5-term poly & 76.3 & 81.92 & 6254 & 0.0413    \\ 
\cite{qian2011uniform} & 6-term poly & 91.6 & 81.92 & 7506 & 0.0355    \\ \hline
\multirow{3}{*}{Ours}&2b BSL  & 645.1 & 0.55 & 342 & 0.0410    \\ 
&4b BSL  & 1290.6 & 0.55 & 710 & 0.0252    \\ 
\rowcolor{lightgray}&8b BSL  & 2581.7 & 0.55 & 1420 & 0.0155    \\ \hline \hline
\end{tabular}    
\vspace{-5pt}
\end{table}


\emph{Softmax Block Comparison}
For the proposed iterative approximate softmax circuit, we selected the design with $B_x=4$ and evaluated different $B_y$s.
We compared our softmax block with the FSM-based softmax design proposed in \cite{hu2018efficient}.
The number of inputs to the softmax function $m$ is set to 64.
As shown in Table~\ref{tab: softmax compare}, 
for the iterative approximate softmax with $B_y=8$, it has reduced the ADP by 1.58$\times$ to 12.6$\times$ with 29.1\% to 22.6\% MAE reduction compared to FSM-based designs. 
Meanwhile, reducing $B_y$ from 8 to 4 can further reduce the ADP by 3.85$\times$ or decrease 44.3\% MAE. 

\begin{table}[!t]
\vspace{-12pt}
\centering
\caption{Compare area, delay, area-delay product (ADP), and mean average error (MAE) 
across different SC circuits for softmax.}
\vspace{-10pt}
\label{tab: softmax compare} 
\setlength{\tabcolsep}{5pt}
\begin{tabular}{cc|cccc}
\hline \hline
\multicolumn{2}{c|}{Design} & \begin{tabular}[c]{@{}c@{}}Area$\downarrow$ \\ (um²)\end{tabular} & \begin{tabular}[c]{@{}c@{}}Delay$\downarrow$ \\ (ns)\end{tabular} & \begin{tabular}[c]{@{}c@{}}ADP$\downarrow$\\(um²·ns)\end{tabular} & MAE$\downarrow$ \\ \hline \hline
\multirow{3}{*}{FSM\cite{hu2018efficient}}& 128b BSL  & 1.26$\times10^{4}$ & 328 & 4.14$\times10^{6}$ & 0.108    \\ 
                        &256b BSL  & 1.26$\times10^{4}$ & 655 & 8.28$\times10^{6}$ & 0.103    \\ 
                        &1024b BSL & 1.26$\times10^{4}$ & 2621 & 3.31$\times10^{7}$ & 0.099   \\ \hline
\multirow{3}{*}{Ours} & $B_y=4$ & 4.23$\times10^{4}$ & 16.12 & 6.81$\times10^{5}$ & 0.106    \\ \rowcolor{lightgray} 
                Ours  & $B_y=8$ & 1.62$\times10^{5}$ & 16.20 & 2.62$\times10^{6}$ & 0.0766    \\ 
                      &$B_y=16$ & 8.73$\times10^{5}$ & 16.28 & 1.42$\times10^{7}$ & 0.0427    \\ \hline \hline
\end{tabular}    
\vspace{-12pt}
\end{table}

\emph{Design Space Exploration}
We explore the design space of the softmax block by varying parameters listed in Table~\ref{tab:paramters} and there are 2916 possible designs in total. 
For $B_x=2$, we found 12 Pareto optimums, 
where ADP varies from $2.45\times 10^5 \mathrm{um^2 \cdot ns} $ to $1.89\times 10^7  \mathrm{um^2 \cdot ns}$ and MAE from $0.0098$ to $0.0714$. 
Similarly, there are 21 Pareto optimums in the design space of $B_x=4$.

The Pareto frontier in the design space illustrates the flexibility of approximate designs in balancing circuit efficiency and computational accuracy.

\begin{figure}[!t]
 \centering
 \includegraphics[width=1\linewidth]{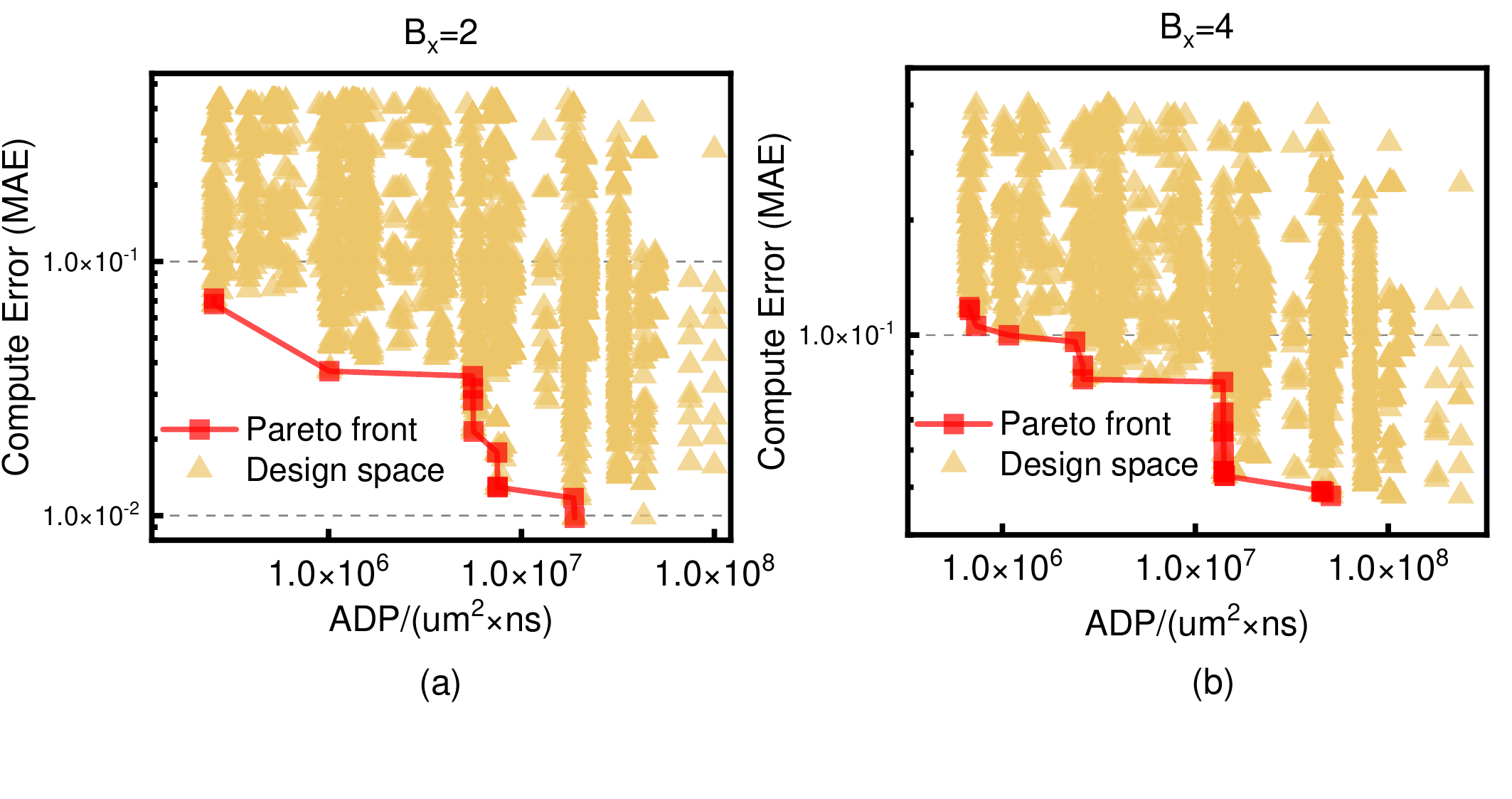}
    \vspace{-30pt}
 \caption{Design space exploration for the iterative approximate softmax block for (a) $B_x=2$ and (b) $B_x=4$.}
 \label{fig:dse}
    \vspace{-10pt}
\end{figure}

\subsubsection{Network Accuracy Comparison}

\begin{table}[!tb]
\centering
\caption{Accuracy Comparison of ViT after using progressive quantization and approximate softmax aware fine-tuning.}
    \vspace{-5pt}
\label{tab:network compare}

\begin{tabular}{c|cc}
\hline \hline
Model                                  & CIFAR10 & CIFAR100 \\ \hline \hline
FP LN-ViT\cite{hassani2021escaping}  & 94.52    & 73.80         \\ 
Baseline low-precision BN-ViT             & 58.13   & 45.76    \\ 
BN-ViT + progressive quant              & 91.12   & 67.16    \\ 
BN-ViT + progressive quant + appr       & 89.27   & 65.36    \\
BN-ViT + progressive quant + appr-aware ft & 90.79   & 66.18    \\ \hline \hline
\end{tabular}
    \vspace{-10pt}
\end{table}

As is shown in Table~\ref{tab:network compare}, the baseline low-precision ViT without using our proposed training pipeline
suffers from a large accuracy degradation even with KD.
Our proposed 
progressive quantization strategy improves the accuracy by 32.99\% and 21.4\% on CIFAR10 and CIFAR100, respectively.
The approximate softmax aware fine-tuning strategy improves the accuracy by 1.52\% and 0.82\% on the two datasets.
With this two-stage training pipeline, our SC-friendly low-precision ViT achieves 32.66\% and 20.42\% higher accuracy than the baseline model on CIFAR10 and CIFAR100, respectively.


\subsubsection{Evaluation of ViT Accelerator}
\begin{savenotes}
\begin{table}[!tb]
    \centering
    \caption{SC accelerator performance with different configurations.} 
    \vspace{-5pt}
    \label{tab: vit comparement}
    
    \begin{threeparttable}
    \setlength{\tabcolsep}{5pt}
    \begin{tabular}{c|cccc}
\hline\hline
Configuration & Softmax &  *Accelerator & \multicolumn{2}{c}{Accuracy} \\
 $[B_y, s_1, s_2, k]$& area $(\rm um^2)$& area  $(\rm um^2)$ & CIFAR10 & CIFAR100 \\
\hline\hline
    $[4,128,2,2]$     &3.15$\times 10^4$  & 4.24$\times 10^6$ & 89.72 & 63.51  \\
    $[8,32,8,3]$      &8.82$\times 10^4$  & 4.47$\times 10^6$ & 90.79 & 66.18   \\
    $[16,128,16,4]$   &4.65$\times 10^5$  & 6.04$\times 10^6$ & 91.07 & 66.63   \\
    $[32,128,16,4]$   &1.16$\times 10^6$  & 8.84$\times 10^6$ & 91.25 & 66.78   \\ 
\hline\hline
 \end{tabular}
    \begin{tablenotes}
\item *In an accelerator, there are $k$ softmax blocks to ensure the fully parallel.
    \end{tablenotes}
\vspace{-10pt}
    \end{threeparttable}
\end{table}
\end{savenotes}

We further conduct accelerator-level evaluations to show the
impact of softmax blocks. Specifically, we select different
softmax block configurations along the Pareto front and
evaluate the area and accuracy impact.
As shown in Table~\ref{tab: vit comparement},
our softmax block only takes a small portion of the total area,
e.g., 1.48\% for small computation BSLs and iterations.
With the increase of BSLs and iterations, although the inference
accuracy increases by more than 1.5\%, the softmax block
area increases drastically by more than 30$\times$,
leading to more than 2$\times$ total area overhead.
Therefore, we might recommend choosing the configuration of $[8, 32, 8, 3]$,
which enables to achieve an accuracy of over 90\% on CIFAR10
with only a marginal increase in total area compared to the
configuration with minimum area.

%


\vspace{-3pt}
\section{Conclusion}
\label{sec:conclusion}
\vspace{-2pt}
In this paper, we investigate the challenges of implementing ViT in SC and propose \method, the
first ViT acceleration with end-to-end SC. 
We propose novel SC circuit blocks for the GELU and the approximate softmax with a parameterized design space and a Pareto optimization. 
The proposed GELU and softmax blocks achieve 56.3\% and 22.6\% error reduction compared to existing SC designs, respectively, and reduce the ADP by 5.29$\times$ and 12.6$\times$, respectively.
We also develop an SC-friendly low-precision ViT
through a two-stage training pipeline including progressive quantization and approximate softmax aware fine-tuning, which significantly improves the accuracy over the baseline low-precision network. 

In summary, the proposed ASCEND is accurate yet efficient. 
Its diverse set of approximation configurations enables flexible adaptation to meet the hardware efficiency and computational accuracy requirements of various applications.
\vspace{-5pt}

\bibliographystyle{IEEEtran}
\bibliography{top_simplified.bib,sc_less.bib}
\end{document}